\documentclass[twoside,fleqn]{article}
\usepackage{espcrc2}
\usepackage{epsfig}
\usepackage{latexsym}

\title{
\vspace{-3.65cm}				
{\normalsize DESY 97--175}	\\[-0.2cm]	
{\normalsize FU-HEP/97--05}	\\[-0.2cm]	
{\normalsize HUB--EP--97/53}	\\[-0.2cm]	
{\normalsize September 1997}	\\		
\vspace{1.7cm}					
Improved Quenched QCD on Large Lattices -- First Results%
\thanks{
Poster presented by D.~Pleiter at Lat97, Edinburgh, U.K.	
}}

\author{M.~G\"ockeler%
           \address{Institut f\"ur Theoretische Physik, Universit\"at
                    Regensburg, D-93040 Regensburg, Germany},
        R.~Horsley%
           \address{Institut f\"ur Physik, Humboldt-Universit\"at zu Berlin,
                    D-10115 Berlin, Germany},
        V.~Linke%
           \address{Institut f\"ur Theoretische Physik,
                    Freie Universit\"at Berlin, D-14195 Berlin, Germany},
        H.~Perlt%
           \address{Institut f\"ur Theoretische Physik, Universit\"at
                    Leipzig, D-04109 Leipzig, Germany},
        D.~Pleiter$^{\rm c,}$
           \hspace{-0.2cm}
           \address{DESY-IfH Zeuthen, D-15735 Zeuthen, Germany},
        P.~E.~L. Rakow$^{\rm e}$,
        G.~Schierholz$^{\rm e,}$
           \hspace{-0.2cm}
           \address{Deutsches Elektronen-Synchrotron DESY,
                    D-22603 Hamburg, Germany},
        A.~Schiller$^{\rm d}$,
        P.~Stephenson$^{\rm e}$
        and
        H.~St\"uben%
           \address{Konrad-Zuse-Zentrum f\"ur Informationstechnik Berlin,
                    D-14195 Berlin, Germany}
}

\begin{document}

\begin{abstract}
Continuing our investigations of quenched QCD with improved fermions
we have started simulations for lattice size $32^3 \times 64$ at
$\beta=6.2$.  We present first results for light hadron masses at
$\kappa=0.13520$, $0.13540$, and $0.13555$.
Moreover we compare our initial experiences on the T3E with
those for APE/Quadrics systems.
\end{abstract}

\maketitle

\section{INTRODUCTION}

High computer costs turned out to be a major problem when performing
quenched QCD simulations at smaller lattice spacing $a$. Improving
the action in order to reduce cut-off effects therefore became an
important goal.

While standard gluonic action has discretization
errors of $O(a^2)$, those for Wilson fermions are of $O(a)$.
Sheikholeslami and Wohlert \cite{sheikholeslami85a} proposed
a modified fermion action:
\begin{eqnarray}
S_F & = & S_F^{(0)} - \\
    &   & \frac{\mbox{i}}{2} \kappa c_{SW}(g^2) a^5
          \sum_x \bar{\psi}(x) \sigma_{\mu\nu} F_{\mu\nu}(x) \psi(x) \nonumber
\end{eqnarray}
where $S_F^{(0)}$ is the standard Wilson action and
\begin{equation}
F_{\mu\nu}(x) = \frac{1}{8 \mbox{i} g a^2}
\sum_{\mu,\nu = \pm} (U(x)_{\mu\nu} - U(x)_{\mu\nu}^{\dagger}).
\end{equation}
If the coefficient $c_{SW}$ of the so-called clover term is chosen
appropriately, this action removes all $O(a)$ errors from on-shell
quantities like hadron masses. A non-perturbative calculation of $c_{SW}$
as a function of $g^2$ was done by the Alpha collaboration \cite{luescher97a}.

Until now the QCDSF collaboration has presented results for the light
hadron spectrum using the improved action for two values of the coupling,
$\beta = 6.0$ and $6.2$
(see \cite{goeckeler97c,stephenson97a}). These calculations
have been carried out on APE Quadrics computers on lattice sizes up
to $24^3 \times 48$. In order to allow a more
reliable estimate of the chiral limit for $\beta = 6.2$
we started calculations with a hopping parameter $\kappa$ closer to the
critical value, $\kappa_c$, on a lattice of size $32^3 \times 64$.

At present we have evaluated $O(75)$ configurations. We hope for
higher statistics, and so the following results should be regarded as
preliminary.

\section{SIMULATION DETAILS}

We perform the quenched QCD simulations at $\beta = 6/g^2 = 6.2$.
To generate a new gauge configuration we use 100 cycles consisting of
a single 3-hit Metropolis sweep followed by 16 over-relaxation sweeps
using the $SU(3)$ algorithm suggested by Creutz \cite{creutz87a}.

We use Jacobi smearing \cite{allton93} for source and sink. We chose
the number of smearing steps to be $N_s = 100$ and for the smearing hopping
parameter we took $\kappa_s = 0.21$, for which the radius of the smeared source
$r a$ is about $3.5a$ which roughly corresponds to $0.4 fm$. Although
we have calculated the propagator for both smeared and unsmeared sink,
we will only use the results for smeared sink here.

The simulations are performed for three different hopping parameters,
$\kappa = 0.13520$, $0.13540$, $0.13555$, with clover coefficient
$c_{SW} = 1.614$ chosen according to \cite{luescher97a}.
For the matrix inversion we mainly use BiCGstab \cite{vorst92,frommer94}.
The minimal residue algorithm is used in case BiCGstab does not
converge. As convergence criterion we chose $r \leq 10^{-15}$,
where $r = |M \chi - \phi| / |\chi|$.

\begin{figure}[bh]
\vspace*{-0.3cm}
\includegraphics[height=5.9cm]{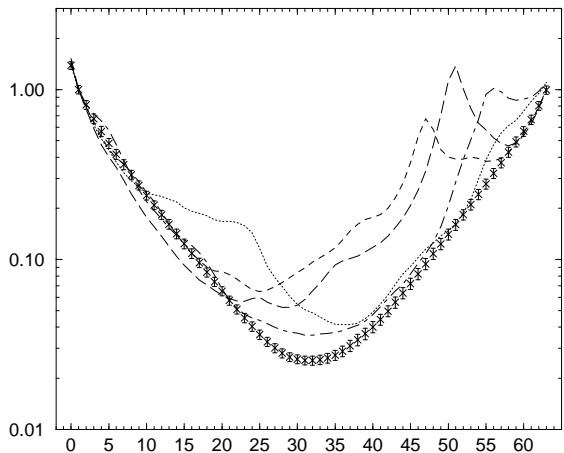}
\vspace*{-0.8cm}
\caption{\footnotesize Pion propagator at $\kappa = 0.13555$ with
         separately plotted exceptional configurations.}
\vspace*{-0.5cm}
\label{fig:except}
\end{figure}

We found up to 4 configurations per $\kappa$ which show an exceptional
pattern (see Fig.~\ref{fig:except}). They have been excluded from
the evaluation.

\section{RESULTS}

\begin{figure}[th]
\vspace*{0.12cm}
\includegraphics[height=8.4cm]{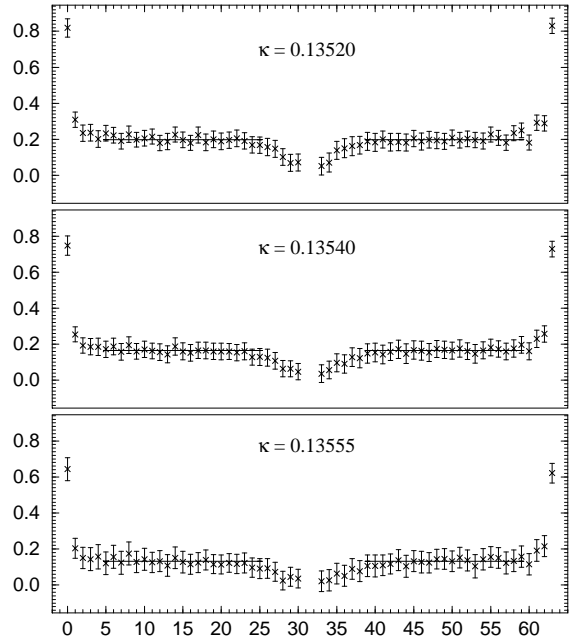}
\vspace*{-0.8cm}
\caption{\footnotesize Effective mass of the Pion.}
\vspace*{-0.5cm}
\label{fig:effmass_pion}
\end{figure}

Until now we have looked at the $\pi$, $\rho$ and nucleon masses.
We find good plateaus when plotting the
effective mass $m(t) = \ln\left[c(t) / c(t+1)\right]$,
as shown in Fig.~\ref{fig:effmass_pion} for the $\pi$.

\begin{figure}[t]
\vspace*{0.12cm}
{\center\includegraphics[height=6.4cm]{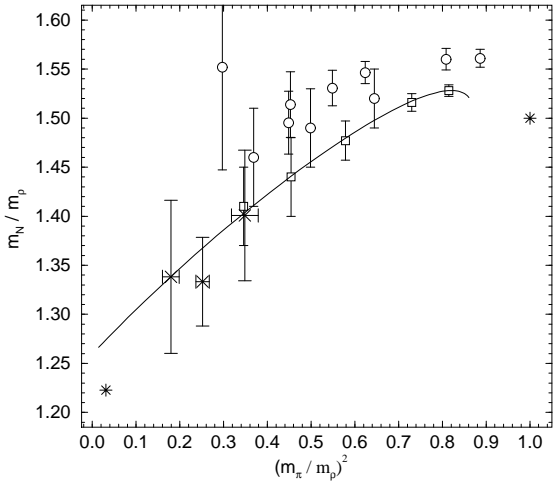}}
\vspace*{-1.0cm}
\caption{\footnotesize
     APE plot at $\beta = 6.2$ for improved ($\Box$ \cite{goeckeler97c})
     and Wilson fermions ($\circ$ \cite{allton97,collins93}). The new
     improved results on larger lattices are marked by $\times$.
     This data can be compared with the mass ratio ($*$) at the
     physical quark mass and in the heavy quark limit.  The solid line
     comes from a fit using the phenomenological ansatz, eq.~\ref{ansatz},
     with the new preliminary data included.}
\vspace*{-0.10cm}
\label{fig:ape}
\end{figure}

Plots of the dimensionless ratio $m_N/m_{\rho}$ as a function of
$(m_{\pi}/m_{\rho})^2$ (so-called APE plot) were found to be rather different
for improved and Wilson fermions at $\beta = 6.0$ \cite{goeckeler97c}.
As can be seen from Fig.~\ref{fig:ape},
the results at $\beta = 6.2$ seem to confirm that the improved results
come closer to the physical value than the Wilson results.

\begin{figure}[t]
\vspace*{0.12cm}
{\center\includegraphics[height=6.4cm]{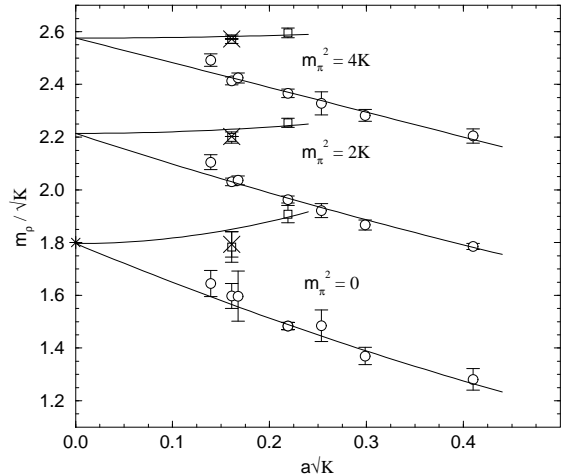}}
\vspace*{-1.0cm}
\caption{\footnotesize The ratio $m_{\rho}/\sqrt(K)$ as a function of the lattice
     spacing for Wilson fermions ($\circ$ \cite{allton97,collins93})
     and improved ($\Box$ \cite{goeckeler97c} and $\times$, this work).
     This is compared with the experimental value
     ($*$) using $\sqrt(K) = 427 MeV$.}
\label{fig:scale}
\vspace*{+0.515cm}
\end{figure}

To see how masses scale as $\beta$ is changed, it has been
suggested \cite{sommer96} that $m_{\rho}$ should be plotted in units of
the square-root of the string tension $K$ which has cut-off errors
of $O(a^2)$ only. In Fig.~\ref{fig:scale} the ratio $m_{\rho}/\sqrt{K}$
is shown as a function of $a \sqrt{K}$ for fixed physical $\pi$ masses
with $m_{\pi}^2 = 0$, $2K$ and $4K$. To obtain the $\rho$ mass
for these values of $m_{\pi}$ we extrapolated, or interpolated,
$m_{\rho}$ using the phenomenological ansatz \cite{goeckeler97c}
\begin{equation}\label{ansatz}
m^2_X = b_0 + b_2 m^2_{\pi} + b_3 m^3_{\pi}, \;\;\; X = \rho, N
\end{equation}

\section{T3E PERFORMANCE}

On the $32^3 \times 64$ lattice our current program needs $13.1 s$
per BiCGstab iteration step on a T3E with 128 DEC Alpha 21164 5/375MHz
microprocessors. Simulations done on a 256 node Quadrics QH2 need $6.3 s$
for the same operations on a $24^3 \times 48$ lattice. Comparing the
peak performance of both machines (T3E: 96 Gflops / QH2: 12.8 Gflops)
one would expect the T3E to do this job about twice as fast as the QH2
(although the calculations on the T3E are done in double precision). Since
the communication overhead in the BiCGstab routines on the T3E is less
then $3\%$, this indicates a single processor performance problem. While
lattice gauge theory applications on the QH2 typically reach sustained
speeds between $30$ and $70\%$ of the peak performance, our T3E code
currently runs at about $10\%$. This might be
explained by two disadvantages of the T3E for this kind of problems: the
number of registers and the cache size of the DEC 21164 microprocessors.
The stream buffers which improve main memory access make the code about
$30\%$ faster.

\section{ACKNOWLEDGMENTS}

The numerical calculations were performed on the CRAY T3E LC 512-512
at the HLRZ J\"ulich and the CRAY T3E LC 128-128 at the
Konrad-Zuse-Zentrum Berlin. We wish to thank the operating staff
for their support.

\end{document}